\documentclass[preprint, 10pt, 3p]{elsarticle}
\usepackage{graphicx}
\usepackage{dcolumn}
\usepackage{bm}
\usepackage{amsthm}
\usepackage{color}
\usepackage{float}
\usepackage{epsfig}
\usepackage{amsfonts}   
\usepackage{amssymb}
\usepackage{amsmath}
\usepackage{hyperref}
\usepackage{cleveref}
\usepackage{cuted}
\usepackage[mathlines]{lineno}

\theoremstyle{definition}
\newtheorem{definition}{Definition}[section]
\newtheorem{theorem}{Theorem}[section]
\newtheorem{lemma}[theorem]{Lemma}
\newtheorem{prop}{Proposition}[section]

\theoremstyle{remark}

\usepackage{scalerel,stackengine}
\stackMath
\newcommand\reallywidehat[1]{%
\savestack{\tmpbox}{\stretchto{%
  \scaleto{%
    \scalerel*[\widthof{\ensuremath{#1}}]{\kern.1pt\mathchar"0362\kern.1pt}%
    {\rule{0ex}{\textheight}}
  }{\textheight}%
}{2.4ex}}%
\stackon[-6.9pt]{#1}{\tmpbox}%
}
\journal{Nuclear Physics A}
\begin{document}

\begin{titlepage}

\centering

\vspace*{2cm}

{\LARGE\bfseries
An Algebra Bundle for Non-Composable Coincidence Probabilities in Gamma-ray Spectroscopy
\par}

\vspace{1.5cm}

{\large
Liam L. Schmidt
\par}

\vspace{0.8cm}

{\normalsize
Department of Physics, University of Guelph\\
50 Stone Road East\\
Guelph, Ontario N1G 2W1, Canada
\par}

\vspace{1.5cm}

{\normalsize
\textbf{Corresponding author:} Liam L. Schmidt\\
Email: \texttt{lschmi04@uoguelph.ca}
\par}

\vfill

\end{titlepage}

\clearpage


\begin{frontmatter}
\begin{abstract}
    Understanding nuclear level schemes from the detection of emitted gamma-rays is, at its core, a combinatorial problem. The transition matrix formalism that treats nuclear decay as a Markov process is a well established treatment for the prediction of emitted and detected gamma-rays. However, with the path algebra being the underlying algebra of such matrix methods, the formalism is restricted to calculating probabilities only for composable paths through the decay. This paper presents a new algebra that allows for the calculations of non-composable coincidence probabilities offering a more complete formalism for the prediction of gamma-ray spectra. The aim of this \textit{coincidence algebra} is to provide a general treatment for predicting all full energy summing peaks for composable and non-composable transitions. Non-composable summing peaks often corrupt  peaks of interest by creating doublets in the energy spectra. The coincidence algebra serves as a tool for investigating any possible doublets produced from full energy coincidence summing. This formalism is tested for the prediction of gamma-ray peaks in the electron capture of $^{133}$Ba and compared to the results a Geant4 simulation of the decay within the GRIFFIN spectrometer. It is also shown how the path algebra formalism may be useful for understanding uncertainty propagation in nuclear level schemes. 
\end{abstract}
\end{frontmatter}

\section{Introduction}
The resultant gamma decay from nuclear decays such as $\beta^+$/$\beta^-$ decay are often depicted by a decay diagram whose structure is analogous to those found in graph theory. Such a diagram is seen in Figure \ref{fig: 1},  where $f_i$ are branching ratios to levels in the daughter nuclear and $x_{ij}$ are the transition probabilities between level $i$ and $j$. This work sets out to further generalize the treatment of these decay diagrams in an algebraic and geometric manner.

An algebraic structure for spectroscopic analysis is proposed that models the gamma decay as a quiver with a defined structure; a quiver is a directed graph upon which a vector space and algebra is composed \cite{Assem_Skowronski_Simson_2006}. It is common practice to treat nuclear levels using a graph theoretical approach. However, the manner in which that is done is through transition matrices \cite{SEMKOW1990437, KORUN1993478} and Markov processes \cite{athanasios}, which take a matrix representation of the more fundamental algebra built from a quiver--the path algebra \cite{Assem_Skowronski_Simson_2006}. The path algebra constructed from a quiver that models a decay scheme allows for probability calculations for coincidence events (as do the transition matrices, though the path algebra keeps cascade terms separate and does not sum with the direct transitions). The conventional path algebra, however, is restricted to consecutive paths; that is, directly connected paths. In this work, the path algebra is extended into a tensor algebra where non-consecutive products may be calculated. This serves as a more general algebraic structure for nuclear level schemes, which may assist in the calculation of coincidence probabilities for gamma-ray spectroscopy. 

\subsection{Transition Matrices}
Conventional treatment for decay diagrams is done with transition matrices \cite{SEMKOW1990437}. General decays can be modeled using strictly lower triangular matrices, where the matrix algebra allows for the calculation of cascade probabilities via powers of the transition matrix. 
\begin{figure}[H]
    \centering
    \includegraphics[width=.7\linewidth]{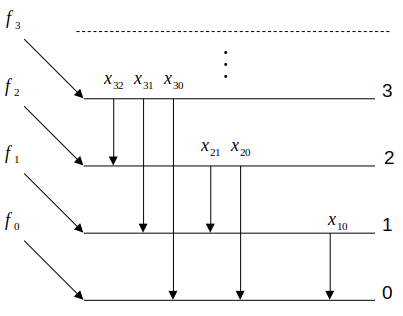}
    \caption{General decay scheme with branching ratios \texorpdfstring{$f_i$}{f_i} and transition probabilities \texorpdfstring{$x_{ij}$}{x_{ij}}.}
    \label{fig: 1}
\end{figure}

\noindent
  For a general decay scheme, as shown in Figure \ref{fig: 1}, the transition probabilities can be stored in the transition matrix 
\begin{equation}
    x = \begin{pmatrix}
        0 &\\
        x_{10} & 0 \\
        x_{20} & x_{21} & 0\\
        \vdots & \vdots & \vdots \\
        x_{n0} &x_{n1} & x_{n2} & \cdots& x_{n n-1} & 0
    \end{pmatrix},
\end{equation}
and the branching ratios are stored in a branching vector
\begin{equation}
    f = (f_0, f_1, f_2, \cdots, f_n).
\end{equation}
Taking powers of the transition matrix will give cascade probabilities on the order of the power in the matrix position of the direct transition. For example, all two gamma cascade probabilities can be calculated from:
\begin{equation}
    x^2 = \begin{pmatrix}
        0 &\\
        0 & 0 \\
        x_{10}x_{21} & 0 & 0\\
        x_{10}x_{31} + x_{32}x_{20} &x_{21}x_{32} & 0  & 0
    \end{pmatrix}.
\end{equation}
To collect all orders of cascade probabilities, from single transitions to $n$ gamma cascades, we may power expand the transition matrix:
\begin{equation}
    \overline{x}^n = \sum_{k=0}^nx^k.
\end{equation}
With the branching vector determining the level at which a given decay starts, the probability that a transition from level $k$ to level $l$ will occur for a given decay can be given as follows:
\begin{equation}
    P_{TM}(x_{kl}) = \left[diag\left(f\overline{x}^n\right)x\right]_{kl}.
\end{equation}
The diagonal matrix $diag\left(f\overline{x}^n\right)$ gives the probability that a given level is reached. To calculate the probability that two transitions, $k\rightarrow l$ and $q\rightarrow m$, occur,  we take the probability of the first transition to occur multiplied by the probability that it leads to the second with
\begin{equation}
    P_{TM}(x_{kl}\cap x_{qm})  = \left[diag\left(\left[f\overline{x}^n\right]_i\right)x\right]_{kl}\left[diag\left(\left[\overline{x}^n\right]_{li}\right)x\right]_{qm}.
\end{equation}
The diagonal matrix $diag\left(\left[\overline{x}^n\right]_{li}\right)$ gives the probability that from level $l$ the decay leads to a given level below it. The above probabilities are probabilities of emission; they indicate that the transition occurs. In gamma-ray spectroscopy we must deal with the probability of detection. It is the map between detected probabilities and emitted probabilities that is the practice of gamma-ray spectroscopy. To model this, the transition matrices can be mapped to include different detection probabilities, such as efficiencies. This is done in \cite{SEMKOW1990437} where observed summing probabilities are given; this paper is motivated by \cite{SEMKOW1990437} where we look to further generalize the mathematical procedure of coincidence probabilities. The work of Semkow et.al. \cite{SEMKOW1990437} first established the transition matrix formalism for nuclear decay schemes, though largely building off the work of \cite{osti_4488528,MCCALLUM1975189}. In their paper, their motivation was to calculate the true coincidence summing corrections that occur in gamma-ray spectroscopy. When detecting the emission of gamma rays in a decay, there is some non-zero probability that two gammas from a cascade can hit your detector making it appear like the detection of a single gamma-ray with energy being the sum of the two. If the direct transition of that cascade exist, then extra counts are added to the peak of the direct transition and counts are lost from the the two smaller energy gamma peaks. When this occurs, we may refer to the loss of counts due to summing as summing out and the gain of counts as summing in. It is crucial in gamma-ray spectroscopy that the coincidence summing effect is well know in order to arrive at an accurate measurement. The transition matrix formalism was extended to include different phenomena such as coincidence summing with X-rays \cite{KORUN1993478}.  The work of \cite{JUTIER20071344} takes a unique approach to modeling coincidence summing with hints of a graph theory, but falls short of reaching the rigorous formalism of quivers and path algebra. Recent work by Schmidt \cite{schmidt2025measurabilitytruecoincidencesumming} gives a more comprehensive history of coincidence summing and also extends the matrix formalism of \cite{SEMKOW1990437} for 180 degree coincidence ( as a correction for coincidence summing) and for gated (coincidence) spectra. The work presented here is a further generalization of the transition matrix formalism for coincidence probabilities in nuclear level schemes.
\par
The transition matrices are limited to creating product probabilities between connected transitions. This is due to the algebra of strictly lower triangular matrices. As shown in \cite{SEMKOW1990437}, this concatenation of probabilities allows for coincidence summing calculations of connected transitions. To calculate the coincidence summing of non-connected transitions, one must extend the algebra. It is the coincidence summing of non-connected transition in particular that motivates this work. To tackle this problem, we propose a structure that offers to deal with coincidence probabilities in the most general sense. The transition matrices emerge out of a more general algebra, called the path algebra of quivers. The transition matrices are isomorphic to the path algebra modded by the ideal of relations that equate all cascades to their direct transition. We extend this path algebra to a \textit{coincidence algebra}. 

\section{Quivers and Path Algebra}
A quiver $\mathcal{Q} = (\mathcal{Q}_0, \mathcal{Q}_1, s,t)$ is a quadruple consisting of two sets: $\mathcal{Q}_0$ (whose elements are called vertices) and $\mathcal{Q}_1$ ( whose elements are called arrows), and two maps $s,t: \mathcal{Q}_1\rightarrow \mathcal{Q}_0$ which associate to each arrow $\alpha \in \mathcal{Q}_1$ its source $s(\alpha) \in \mathcal{Q}_0$ and its target $t(\alpha) \in \mathcal{Q}_0$, respectively. A \textit{path} of length $l$ is a sequence of connected arrows from $a\in\mathcal{Q}_0$ to $b\in\mathcal{Q}_0$ written as follows:
 \begin{equation}
     (a|\alpha_1,...\alpha_l|b),
 \end{equation}
 where $\alpha_i \in \mathcal{D}_1$ is an arrow from vertex $a$. A path of length $l=0$ is called the stationary path, denoted by
 \begin{equation}
     \epsilon_a = (a||a)
 \end{equation}
 for vertex $a$. The path algebra $k\mathcal{Q}$ of $\mathcal{Q}$ is the k-algebra whose underlying k-vector space has as its basis the set $\mathcal{P}$ of all paths $(a|\alpha_1, ...\alpha_l|b)$ of length $l\geq 0$ in $\mathcal{Q}$ and such that the product of two basis vectors $(a|\alpha_1, ...\alpha_l|b)$ and $(c|\beta_1, ...\beta_k|b)$ of $k\mathcal{Q}$ is defined by
    \begin{equation}
        (a|\alpha_1, ...\alpha_l|b)\cdot(c|\beta_1, ...\beta_k|d)=\delta_{bc}(a|\alpha_1, ...\alpha_l,\beta_1, ...\beta_k|d).
    \end{equation}
The product of basis elements is extended to arbitrary elements of $k\mathcal{Q}$ by distributivity. For the sake of brevity, we use a hat notation to denote paths where for arbitrary paths we write $\widehat{p}$ and for distinct paths the concatenation of arrows, $\widehat{\alpha_1...\alpha_l}$, will be made explicit where the source and targets can be given by the $s$ and $t$ maps. The stationary paths are a complete set of idempotents for the algebra such that their sum gives the global identity element:
\begin{equation}
\mathbb{I}_\mathcal{Q} = \sum_{a\in \mathcal{Q}_0}\widehat{\epsilon}_a.
\end{equation}
The vector space $k\mathcal{Q}$ has a direct sum decomposition into paths of different lengths:
\begin{equation}
    k\mathcal{Q
    } = k\mathcal{Q}_0 \oplus k\mathcal{Q}_1\oplus ...\oplus k\mathcal{Q}_n
\end{equation}
where $k\mathcal{Q}_l$ is the subspace whose basis $\mathcal{P}_l$ is the set of all paths of length $l$. The arrow ideal $R_\mathcal{D}$ , which is defined as isomorphic to the quotient vector space $k\mathcal{Q}/k\mathcal{Q}_0$, may be decomposed into \textit{source subspaces} $\mathcal{S}_i$  or target subspaces  $\mathcal{T}_i$ composed of basis vectors with their source or target as level $i$, respectively:
\begin{equation}
    R_\mathcal{D} = \mathcal{S}_1 \oplus \mathcal{S}_2\oplus ...\mathcal{S}_n,
\end{equation}
\begin{equation}
    R_\mathcal{D} = \mathcal{T}_0 \oplus \mathcal{T}_1\oplus ...\mathcal{T}_{n-1}.
\end{equation}
We now define the following bilinear forms.
\begin{definition}
    For two paths $\widehat{p_i},\widehat{p_j}\in P$, the \textit{source bilinear form}$\langle,\rangle_s:k\mathcal{D}\rightarrow\mathbb{R}$ is given by
    \begin{equation}
        \langle \widehat{p_i},\widehat{p_j}\rangle_s = \delta_{s(\widehat{p_i})s(\widehat{p_j})}.
    \end{equation}
    Two paths $\widehat{p_i},\widehat{p_j}$ are \textit{diverging paths} if $\langle \widehat{p_i},\widehat{p_j}\rangle_s$ = 1
\end{definition}
\begin{definition}
    For two paths $\widehat{p_i},\widehat{p_j}\in P$, the \textit{target bilinear form}$\langle,\rangle_t:k\mathcal{D}\rightarrow\mathbb{R}$ is given by
    \begin{equation}
        \langle \widehat{p_i},\widehat{p_j}\rangle_t = \delta_{t(\widehat{p_i})t(\widehat{p_j})}.
    \end{equation}
    Two paths $\widehat{p_i},\widehat{p_j}$ are \textit{converging paths} if $\langle \widehat{p_i},\widehat{p_j}\rangle_t$ = 1
\end{definition}
\begin{definition}
    For two paths $\widehat{p_i},\widehat{p_j}\in P$, the \textit{path bilinear form}$\langle,\rangle_t:k\mathcal{D}\rightarrow\mathbb{R}$ is given by
    \begin{equation}
        \langle \widehat{p_i},\widehat{p_j}\rangle_p = \delta_{s(\widehat{p_i})s(\widehat{p_j})}\delta_{t(\widehat{p_i})t(\widehat{p_j})}.
    \end{equation}
    Two paths $\widehat{p_i},\widehat{p_j}$ are \textit{equivalent paths} if $\langle \widehat{p_i},\widehat{p_j}\rangle_t$ = 1
\end{definition}

These bilinear forms extend to the entire vector space with distributivity. Projectors into the different subspaces can now be defined:
\begin{definition}
    The \textit{source projector} $S_i:k\mathcal{Q}\rightarrow\mathcal{S}_i$ projects a  vector $d \in k\mathcal{Q}$ into its source subspace $\mathcal{S}_i$:
    \begin{equation}
        S_i(d) = \sum_{\widehat{p}\in\mathcal{P}_1} \delta_{s(\widehat{p})i}\langle \widehat{p},d \rangle_s \widehat{p}.
    \end{equation}
\end{definition}
\begin{definition}
    The \textit{target projector} $T_i:k\mathcal{Q}\rightarrow\mathcal{T}_i$ projects a vector $d \in k\mathcal{Q}$ into its target subspace $\mathcal{T}_i$:
    \begin{equation}
        T_i(d) = \sum_{\widehat{p}\in\mathcal{P}_1} \delta_{t(\widehat{p})i}\langle \widehat{p},d \rangle_t \widehat{p}.
    \end{equation}
\end{definition}
\begin{definition}
    The \textit{target vertex projector} $V_t:k\mathcal{Q}\rightarrow k\mathcal{Q}_0$ projects a vector $d\in k\mathcal{D}$ into its idempotent subspace in a manner that sums over \textit{converging} paths:
    \begin{equation}
        V_t(d) = \sum_{\widehat{\epsilon}\in\mathcal{P}_0}\langle\widehat{\epsilon}, d\rangle_t\widehat{\epsilon}.
    \end{equation}
\end{definition}
\begin{definition}
    The \textit{source vertex projector} $V_s:k\mathcal{Q}\rightarrow k\mathcal{Q}_0$ projects a vector $d\in K\mathcal{D}$ into its idempotent subspace in a manner that sums over \textit{diverging} paths:
    \begin{equation}
        V_s(d) = \sum_{\widehat{\epsilon}\in\mathcal{P}_0}\langle\widehat{\epsilon}, d\rangle_s\widehat{\epsilon}.
    \end{equation}
\end{definition}
\begin{definition}
    The \textit{branching projector} $B:k\mathcal{Q}\rightarrow k\mathcal{Q}_0$ projects a decay vector $d \in k\mathcal{D}$ into its idempotent subspace $K\mathcal{D}_0$:
    \begin{equation}
        B(d) = \sum_{\widehat{\epsilon}\in\mathcal{P}_0} \langle \widehat{\epsilon},d \rangle_p \widehat{\epsilon}.
    \end{equation}
\end{definition}
The above definitions will aid in the following investigation in nuclear decay schemes formalized with quivers.
\subsection{Decay Quiver}
\begin{definition}
    A \textit{decay quiver} is a finite connected acyclic quiver $\mathcal{D}=((\mathcal{D}_0,\preceq), \mathcal{D}_1, s, t)$ where $\mathcal{D}_0$ is the ordered set of vertices (levels), $\mathcal{D}_1$ is the set of arrows (transitions/gammas), and $s,t:\mathcal{D}_1\rightarrow \mathcal{D}_0$ are maps that associate to each arrow in $\mathcal{D}_1$ its \textit{source} and \textit{target}, respectively. 
\end{definition}
We define a possible relation on $D_0$:
\begin{definition} 
    Let $N:\mathcal{D}_0\rightarrow \mathbb{N}$  and $\widehat{\epsilon}\in \mathcal{D}_0$ where 
    \begin{equation}
        N(\epsilon) = \sum_{\widehat{p}\in\mathcal{P}_1}\langle\widehat{\epsilon}, \widehat{p}\rangle_s,
    \end{equation}
that is, it returns the number of diverging arrows for that vertex. 
\end{definition}
\begin{prop}
    A decay quiver with n levels is fully connected if a total ordering can be defined on the set of vertices $\mathcal{D}_0$ via the relation $\preceq$ with a bijective map $N:\mathcal{D}_0\rightarrow \mathbb{N}_{< n}$ defined by
     \begin{equation}
        \widehat{\epsilon}_a \preceq \widehat{\epsilon}_b \Leftrightarrow N(\widehat{\epsilon}_a) \leq N(\widehat{\epsilon}_b).
    \end{equation}
\end{prop}
In all further derivations and discussion, we take any decay quiver with n levels to be fully connected unless otherwise stated. The bijection between $\mathcal{D}_0$ and $\{0,1,2,...n-1\}$ gives the number of arrows in the quiver, that is, the cardinality of the arrow set, to be $|\mathcal{D}|= n(n-1)/2$. Any given decay vector $d\in K\mathcal{D}$ can be decomposed into a sum of its source vectors (or target vectors) $d_i$, which are the vectors of the source(target) projectors, and its branching vector, which is the projected vector of stationary paths: 
    \begin{equation}\label{eqn:eqn24}
        d = B(d) + \sum_{i=0}^n S_i(d)=  b + \tau.
    \end{equation}
The source vectors $S_i(d)$ and the branching vector $b$ are then unit vectors under the $l_1$-norm $||.||$, given the probability conditions of the level transitions and the branching. The path algebra vector space $k\mathcal{D}$ is equipped with the norm topology induced by $l_1$, which is equivalent to the standard finite-dimensional topology. Decay vectors may thus be classified as unit vectors under a \textit{decay norm} $||d||_\mathcal{D}$ defined as
\begin{equation}
     ||d||_\mathcal{D} = \frac{1}{n} \sum _{j<i}^n|x_{ij}|=1,
\end{equation} 
which is a sum over all the components of the decay vector divided by the number of levels. This norm further defines the subset in which the decay vectors live in; that is, in the subset $k\mathcal{D}_0\oplus k\mathcal{D}_1$ with the constraint $||d||_\mathcal{D} = 1$. 
Analogous to the probabilities given in equations 5 and 6, we can formulate the probability of a given transition in the path algebra using the projectors we defined:
\begin{equation}\label{eqn:eqn26}
        P(\widehat{p}_i) = \left\langle V_t\left(b\cdot \overline{\tau}^n \right) \cdot \tau, \widehat{p_i}\right\rangle_p,
\end{equation}
along with the coincidence probability,
\begin{align}
        P(\widehat{p}_i\cap\widehat{p}_j) = \left\langle V_t\left(b\cdot \overline{\tau}^n \right) \cdot \tau, \widehat{p_i}\right\rangle_p\langle V_t(\epsilon_{t(\widehat{p}_i)}\cdot\overline{\tau}^n) \cdot \tau, \widehat{p}_j\rangle_p,
\end{align}

where $\tau = d - b$ as defined in Eq. \ref{eqn:eqn24}, that is, the decay vector with the branching terms subtracted off. In this formulation, the target vertex projector acts as the diagonalization (left multiplying) of the decay vector with the branching from above terms multiplied through. To select a given transition the path equivalence bilinear form is used. It should be noted, given that the target vertex projector gives back a vector of stationary paths, when multiplied onto the decay vector $d$, what is given is a vector in $k\mathcal{D}_1$ still. Therefore, when finding a given path with the path form only single terms are produced-- no equivalence between different path lengths is produced yet, as would be given by the path form in general. To arrive at notions of coincidence summing, which in this formalism we may generally state as the path equivalence between paths of different lengths, one must power expand the decay vector which $V_t$ is being multiplied onto to create the higher order path terms. Algebraically, one may think of coincidence summing as path equivalence; however, a more complete depiction of coincidence summing is more aptly posed in the context of detection maps. 

\subsection{Detection Maps}
For a given decay quiver, a decay vector living in the path algebra models a nuclear level scheme with n levels. The decay vectors components give the probability for the transition corresponding to the basis vector path. As such, the decay vectors are units vectors under the decay norm. However, there of course exist vectors in the path algebra that do not correspond to transition probabilities satisfying the norm condition, but may still be considered as probabilities. In this manner, detection vectors may also live in the path algebra where the components not only include the transition probabilities but also efficiencies or different detection probabilities. We posit then that the path algebra for a given quiver is the space of all detection nuclear level schemes, such that transition probabilities are mapped to detection probabilities in this space.
If we compute probabilities via Eq. \ref{eqn:eqn26} with a detection mapped decay vector, where efficiencies are now multiplied onto the transition probabilities, what we get is not the probability of emission but the probability of detecting that gamma/transition. In general, for more realistic observations of decaying nuclei, there are other factors beside detector efficiency that go into the detected probabilities that one doing gamma spectroscopy must consider when attempting to map their way back to the transition probabilities; one can consider a number of different detection maps that take the decay vectors describing nuclear decay schemes into different vectors that account for different probabilities involved in detection. A detection map that corresponds to the full energy hit of a gamma may be given by,
\begin{equation}
    \varphi_h(\tau) = \sum_{\widehat{p} \in \mathcal{P}_1}e^p_{\widehat{p}}\langle\widehat{p},\tau \rangle_p \widehat{p},
\end{equation}
where $e^p_{\widehat{p}}$ is the peak efficiency for the gamma associated to the path $\widehat{p}$ of length one. We may also consider the avoidance of a gamma at any energy ( full detection or Compton scattered). To do so, however, we must consider the detector geometry. When talking about gammas summing together and avoiding summing out one must consider angular correlations: the probability that two gammas are emitted at a certain angle. For now, we just consider isotropic emission such that the probability for two gammas hitting the same detector is just given by the number of detectors. Therefore, we may map a decay vector to a detection vector that accounts for the probability of it hitting a given detector at any energy:
\begin{equation}
    \varphi_a(\tau) = \sum_{\widehat{p} \in \mathcal{P}_1}\frac{e^t_{\widehat{p}}}{N}\langle\widehat{p},\tau \rangle_p \widehat{p},
\end{equation}
for $N$ detectors ( one may imagine N detectors spherically placed around an emission source) where $e^t_{\widehat{p}}$ is the total efficiency. To arrive at a notion of summing out, we consider the probability that a given gamma does \textit{not} hit a given detector at any energy:
\begin{equation}
    \varphi_o(\tau) = \sum_{\widehat{p} \in \mathcal{P}_1}\left[ 1-\frac{e^t_{\widehat{p}}}{N}\right]\langle\widehat{p},\tau \rangle_p\widehat{p},
\end{equation}
These are all maps analogous to those defined in \cite{SEMKOW1990437} with the transition matrix formalism and used in the literature afterwards to discuss summing corrections. To account for the summing in, cascade probabilities are built via the power expansion of the decay vector. However, if the power expansion creates the gammas of interest with cascade terms to account for summing in, we must now consider the detector geometry involved for the summing in. Maintaining our isotropic approximation, summing in probabilities can just be considered with the number of detectors (\cite{SEMKOW1990437} considers only one detector so the efficiencies are sufficient in accounting for summing probabilities. For multiple detectors one must extend this of course). We then redefine the power expansion used for detection mapped decay vectors with isotropic emission.
\begin{equation}
    \overline{\varphi(\tau)}^{n}_N = \sum_{k}^n\frac{\tau^k}{N^{k-1}},
\end{equation}
 with $k$ the degree of summing. We may now give a path algebraic formulation for the probability vector that provides the probability of detection for a given transition where summing in and summing out effects are considered:\footnote{This is analogous to the matrix $S=NAM$ given in \cite{SEMKOW1990437} with the rate $R$ taken out}.

\begin{equation}\label{eqn:eqn32}
\Gamma(d) = V_t\left(b\cdot \overline{\varphi_o(\tau)}^n\right)  \cdot \overline{\varphi_h(\tau)}^n_N\cdot V_s\left(\overline{\varphi_o(\tau)}^n\right).
\end{equation}
In contrast from equation 26, the target vertex projector now contains probabilities multiplied onto vertices (stationary paths) that account for summing out from above, $\overline{\varphi_h(\tau)}^n_N$ contains the paths with summing in given via the power expansion, and the source vertex projector now gives summing out from below. All of which, when given together as shown above, provide a vector where the probability of detecting a transition is given by the component of the corresponding basis vector/path. However, different from the transition matrix formalism where the summing in effects are added with the direct transition, the summing in terms exist in separate paths, i.e. they have their own basis vector. To add them back together, the path equivalence form, which equates paths with the same source and target--but does not distinguish between path length, can be used to calculate probabilities for detecting a given transition with summing effects accounted for:
\begin{equation}
    P_{PA}(\widehat{p}_i) = \langle \Gamma(d),\widehat{p}_i \rangle _p.
\end{equation}
Up to here, this new formulation just restates the transition matrix formalism with the caveat that summing in terms are kept separate; the path equivalent quotient path algebra, that is, the path algebra with the path equivalence modded out, is isomorphic to the transition matrices. However, this new formalism may now be extended beyond the path concatenation that we are restricted to with the transition matrices and also allow for a greater separability between equivalent paths.  If we wish to allow for not only paths -- being connected arrows --  but pairs, or even sets, of non-connected arrows (i.e., coincidences in general) in the algebra (as shown in Figure 2), we must extend the algebra to a tensor decay space defined in the following section 3.

\begin{figure}[H]
    \centering
    \includegraphics[width=.7\linewidth]{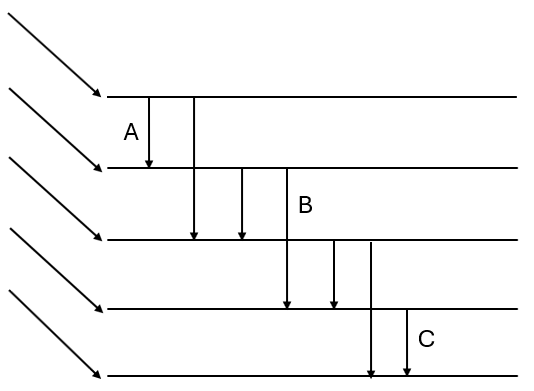}
    \caption{General decay diagram showing the connected transitions A and B and the non-connected or \textit{coincidence} transitions A and C.}
    \label{fig:placeholder}
\end{figure}

\section{Coincidence Algebra Bundle}
\subsection{Coincidence Tensor Space}
The path algebra vector space can be extended to include basis vectors that afford product probabilities between non-connected paths by taking its tensor product expansion. The \textit{tensor decay space} $\mathcal{T}$ can be defined as follows:
\begin{equation}
    \mathcal{T} = k\mathcal{D} \oplus \left( k\mathcal{D} \otimes k\mathcal{D}\right)\oplus...\oplus\left(k\mathcal{D}^{\otimes^n}\right) = \bigoplus_{i=0}^n k\mathcal{D}^{\otimes^i}
\end{equation}
This larger space contains --  as basis vectors --  all combinations of paths in the form of tensor products; not only non-connected but overlapping paths as well. However, it is not this total space that we require for our coincidences. In fact, only the physically realized coincidences are needed--these being the pairs of arrows that can be observed in a single decay event. For this reason, we consider a subset of this tensor space:
\begin{definition}
\label{def:ordering}
    Let $\mathbb{P}(\mathcal{P}_0\cup\mathcal{P}_1)$ be the power set of $\mathcal{P}_0\cup\mathcal{P}_1$, that is, the set of all possible subsets of $\mathcal{P}_0\cup\mathcal{P}_1$. The coincidence set $\mathcal{O}$ is a subset of $\mathbb{P}(\mathcal{P}_0\cup\mathcal{P}_1)$ and is defined as
        \begin{align}
         \mathcal{O}\subset \mathbb{P}(\mathcal{P}_0\cup\mathcal{P}_1) = \left\{ u \in \mathbb{P}(\mathcal{P}_0\cup\mathcal{P}_1)  \Big| |u|\geq 2, \forall \widehat{p}_1, \widehat{p}_2 \in u,
          s(\widehat{p}_2)\preceq t(\widehat{p}_1) \vee s(\widehat{p}_1)\preceq t(\widehat{p}_2) \right\},
    \end{align}
    where the paths in a given element of $\mathbb{P}(\mathcal{P}_0\cup\mathcal{P}_1)$ with cardinality greater than or equal to two do not overlap. 
\end{definition}
The coincidence set may now be used to define the coincidence space.
\begin{definition}\label{def:coinspace}
    Let $o\in \mathcal{O}$ be totally ordered such that $\forall \widehat{p}_1, \widehat{p}_2 \in o, \widehat{p}_1 < \widehat{p}_2 \iff  N(s(\widehat{p}_2)) < N(s(\widehat{p}_1))$; that is, it is ordered by the higher paths in the quiver,  and let $o$ be indexed by this ordering. The \textit{coincidence vector space} $ \mathcal{C}$ is defined as
    \begin{equation}
        \mathcal{C}:= span\left\{t=\bigotimes_{i=1}^{|o|}\widehat{p}_i \Big| \widehat{p}_i \in o, \forall o \in \mathcal{O}\cup\mathcal{P}_0\cup\mathcal{P}_1\right\}/\mathcal{I}.
    \end{equation}
    Where $\mathcal{I}$ is an identity ideal of relations which allow for the stationary paths to act as local identities still:
    \begin{equation}
        \mathcal{I} = \langle s(\widehat{p}_i)\otimes\widehat{p}_i - \widehat{p_i},\widehat{p}_i\otimes t(\widehat{p}_i )- \widehat{p_i}| \forall \widehat{p}_i\in \mathcal{P}_1\rangle.
    \end{equation}
    We include $\mathcal{P}_0\cup\mathcal{P}_1$ to retain the single transition that were exclude in the definition of $\mathcal{O}$. The basis elements of$\mathcal{C}$ are called \textit{coincidences}.
\end{definition}
\begin{definition}
\label{def:coingrading}
    The coincidence vector space $\mathcal{C}$ is graded by the number of
    transitions in a tensor product. That is,
    \begin{equation}
        \mathcal{C}=\bigoplus_{i=0}^n \mathcal{C}_i,
    \end{equation}
    where $\mathcal{C}_i$ is the subspace spanned by basis elements
    containing exactly $i$ transition paths, with stationary paths assigned
    degree zero.\footnote{Note, the grading on $\mathcal{C}$ is not the same grading that is on the tensor space $\mathcal{T}$ as tensor degree grading is broken by the quotient with the identity ideal. However, with grading defined by degree of coincidence with stationary paths degree zero and transitions degree one, the grading of the coincidence space does follow.}
\end{definition}

\begin{definition}
\label{def:cointsmaps}
    The source and target maps defined in the path algebra extend to coincidences (tensor products) as follows:
    \begin{equation}
        s(\widehat{p}_i\otimes\widehat{p}_j\otimes...\otimes\widehat{p}_n) = s(\widehat{p}_i)
    \end{equation}
    \begin{equation}
        t(\widehat{p}_i\otimes\widehat{p}_j\otimes...\otimes\widehat{p}_n) = t(\widehat{p}_n)
    \end{equation}
\end{definition}

\subsection{Coincidence Algebra Bundle}

With the coincidence decay space now defined, we may consider an algebra on this vector space. The path algebra defined for a quiver uses concatenation of paths. In this manner, a given vector, when squared,  will acquire components of concatenated paths where the coefficients are multiplied (i.e., it is a graded algebra). In the context of decays this alludes to coincidence probabilities or joint probabilities of the two paths. When considering non-connected pairs/sets-of-transitions, i.e. coincidences, one must consider what type of multiplication is appropriate when building them out of transitions. In the event that the coincidences contain directly connected transitions, it must also reduce to the original path algebra. When considering coincidences and their probabilities, what matters is precisely how they are not connected. A degree of connectivity can be defined by a sum over connecting paths. This is made clear by considering two gammas in the decay. If one gamma is at a higher level, what determines if that decay will lead to the second gamma is the sum of the probabilities contained in the different paths that connect them. This can be seen in Figure 2 where gamma A and C are only partially connected given that there a exist a decay from A to the ground state that does not emit gamma C. The connectivity between A and C is the probability that the tail vertex of A leads to the source vertex of C given the intermediary transitions.

\par
If the two gammas \textit{are directly connected}, i.e., concatenate to a larger path, then the sum will simply reduce to one. In the path algebra formalism, the decay scheme exist as a decay vector in the decay vector space; we simply get correct product of probabilities since the transition probabilities are contained in the components of the decay vector, i.e., they are inherently in the vector multiplication by use of the decay vector. When a decay vector existing in the decay vector space -- which is now taken to be a subspace of the coincidence decay space -- is squared, the product needs to allow for coincidence terms where the coefficients have the appropriate values; i.e., the sum of the paths connecting them. The vector multiplication in the coincidence decay space therefore must depend on the decay vector; i.e., the product of two basis vectors in the coincidence space needs to give a third vector that has a scalar factor which is a function of the decay vector. An single algebra therefore can not be defined on the entire decay vector space and what is needed is a family of algebras. In particular, what we arrive at is an \textit{algebra bundle}, i.e., a fiber bundle where the base space is the path algebra of the decay quiver  and the fibers are a coincidence algebra defined at every point in the base space. Before we define the algebra, let us start with the notion of connectivity between two paths.

\begin{definition}
    The \textit{scalar connection} $c_d:k\mathcal{D}\times k \mathcal{D}\rightarrow \mathbb{R}$ between two paths $\widehat{p_i},\widehat{p_j}\in \mathcal{P}$ in a decay vector $d$ is defined as follows:
    \begin{equation}
        c_d(\widehat{p_i},\widehat{p_j}) =  \left\langle V_t\left(t(\widehat{p}_i)\cdot\overline{\tau}^n\right), s(\widehat{p}_j)\right\rangle_p.
    \end{equation}
    Similar to the path algebra in Eq. (9), the product of basis elements (paths in $\mathcal{P}$), extends to arbitrary elements of $k\mathcal{D}$ by distributivity, allowing the scalar connection to be defined on $k\mathcal{D}\times k \mathcal{D}$. 
\end{definition} 
\begin{lemma}
\label{lemma:cocycle}
    The scalar connection satisfies the 2-cocycle condition
    \begin{equation}
        c_d(\widehat{p_j},\widehat{p_k})c_d(\widehat{p_i},\widehat{p_j}\otimes\widehat{p_k})=c_d(\widehat{p_i},\widehat{p_j})c_d(\widehat{p_i}\otimes\widehat{p_j},\widehat{p_k}).
    \end{equation}
\end{lemma}
\begin{proof}
    By Definition~\ref{def:cointsmaps}, the scalar connection only depends on the source and target maps of the paths or coincidences. Consequently,
    \begin{equation}
        c_d(\widehat{p_i},\widehat{p_j}\otimes\widehat{p_k}) = c_d(\widehat{p_i},\widehat{p_j}),
    \end{equation}
    likewise
    \begin{equation}
        c_d(\widehat{p_i}\otimes\widehat{p_j},\widehat{p_k}) = c_d(\widehat{p_j},\widehat{p_k}).
    \end{equation}
    Thus, both sides of Eq. 41 reduce to $c_d(\widehat{p_i},\widehat{p_j})c_d(\widehat{p_j},\widehat{p_k})$.
    
\end{proof}
As previously mentioned, when the paths are connected by a single vertex, the scalar connection just returns a value of one. In separated cases, the scalar connection gives a sum over paths with the weights given by the decay vector. In the meaning  of the decay, the scalar connection gives the probability that the higher transition will lead to the lower transition. If the scalar connection is one, the two paths are \textit{directly connected }. If the scalar connection is less than one, the two paths are  \textit{partially connected }. With the scalar connection, we may now define the coincidence algebra.
\begin{definition}
    \label{def:coin_algebra}
    Let $d \in k\mathcal{D}$ be a decay vector. The coincidence algebra
    $\mathcal{C}_d$ is the vector space $\mathcal{C}$ equipped with the
    bilinear multiplication $\cdot_d$ defined on basis elements by
    \begin{equation}
        \widehat{p_i}\cdot_d \widehat{p_j}
        =
        c_d(\widehat{p_i},\widehat{p_j})
        \widehat{p_i}\otimes\widehat{p_j},
    \end{equation}
    for $\widehat{p_i},\widehat{p_j}\in k\mathcal{D}$, and extended
    linearly to $\mathcal{C}$ via Definition~\ref{def:cointsmaps}. 
\end{definition}
\begin{prop}
    The multiplication $\cdot_d$ defined in Definition~\ref{def:coin_algebra}
    is closed and associative on $\mathcal{C}$. Hence, $\mathcal{C}_d$ is an algebra.
\end{prop}
\begin{proof}
    Closure follows given that the product of every basis element either give a coincidence basis element following the grading in Definition~\ref{def:coingrading} or is zero due to ordering constraints of transitions shown in Definition~\ref{def:ordering}. Associativity is shown as follows. Given $\widehat{p_i},\widehat{p_j},\widehat{p_k}\in C$,
    \begin{eqnarray}
        (\widehat{p_i}\cdot_d \widehat{p_j}
        )\cdot_d \widehat{p_k} &=& c_d(\widehat{p_i},\widehat{p_j})\widehat{p_i}\otimes\widehat{p_j}\cdot_d  \widehat{p_k}\\\nonumber
        &=&c_d(\widehat{p_i},\widehat{p_j})c_d(\widehat{p_i}\otimes\widehat{p_j},\widehat{p_k})\widehat{p_i}\otimes\widehat{p_j}\otimes\widehat{p_k}\nonumber
    \end{eqnarray}\
    and
    \begin{eqnarray}
        \widehat{p_i}\cdot_d (\widehat{p_j}
        \cdot_d \widehat{p_k}) &=& c_d(\widehat{p_j},\widehat{p_k})\widehat{p_i}\cdot_d \widehat{p_j} \otimes\widehat{p_k}\\\nonumber
        &=&c_d(\widehat{p_j},\widehat{p_k})c_d(\widehat{p_i},\widehat{p_j}\otimes\widehat{p_k})\widehat{p_i}\otimes\widehat{p_j}\otimes\widehat{p_k}.\nonumber
    \end{eqnarray}
    Where the above equations are made equal by Lemma~\ref{lemma:cocycle}. The multiplication is well defined on the quotient by use of Definition~\ref{def:cointsmaps}. Any coincidences inputted into the scalar connection are done so via their source and target projections. Therefore 
    \begin{equation}
        c_d(s(\widehat{p}_i)\otimes\widehat{p}_i\otimes t(\widehat{p}_i), s(\widehat{p}_j)\otimes\widehat{p}_j\otimes t(\widehat{p}_j)) =  c_d(s(\widehat{p}_i, \widehat{p}_j) 
    \end{equation}
 
\end{proof}
Moving forward, we must consider consistent decay structures in the decay vector space. If a decay vector space is constructed from a fully connected decay quiver, then the space will allow for all possible transitions. However, for a given gamma decay scheme, it is usually not the case that it can be represented by a fully connected decay quiver. Rather, its decay structure exist in some subspace of the fully connected decay vector space where forbidden transitions are dimensions excluded in the subspace describing the decay structure. We must recognize this because when comparing coincidence algebras at different decay vectors $d\in k\mathcal{D}$, if the decay vectors are of a fundamentally different decay structure, that is, one includes a transition that is forbidden in the other, then the map from one decay vector to another may give a scalar connection of zero--not due to failure of transitions being properly ordered, but due to transitions probabilities being equal to zero. If we wish to establish any equivalence between coincidence algebras at different decay vectors, we must restrict the equivalence to structurally consistent decay schemes, which we define as follows. 
\begin{definition}
\label{def:consistent_decay_structure}
Let
\[
d=\sum_{p\in\mathcal{P}_0\cup\mathcal{P}_1}x_p \widehat{p}
\]
be a decay vector. The support of $d$ is defined by
\[
\operatorname{supp}(d)
=
\{p\in\mathcal{P}_0\cup\mathcal P_1\mid x_p\neq0\}.
\]
Two decay vectors $d,d'\in k\mathcal D$ are said to have the same
decay structure, written
\[
d\sim d',
\]
if
\[
\operatorname{supp}(d)=\operatorname{supp}(d').
\]
The corresponding consistent decay region associated with $d$ is the
equivalence class
\begin{equation}
B_d=[d]=\{d'\in k\mathcal D\mid d'\sim d\}.
\end{equation}

\begin{lemma}
For paths $\widehat{p}_i$ and $\widehat{p}j$, the scalar connection $c_d(\widehat{p}_i,\widehat{p}_j)$ defines a continuous function on the structural equivalence class
$B_d$, where $B_d$ is equipped with the subspace topology inherited from $k\mathcal{D}$.
\end{lemma}
\begin{proof}
    The scalar connection $c_d$ is constructed from projections of products of vectors in $B_d$ and is therefore restricted to polynomials of positive non-zero coefficients which is continuous for all $d \in B_d$ .
\end{proof}
\begin{lemma}
    \label{lemma:isomorphisms}
    For two coincidence algebras $C_d$ and $C_{d_0}$ with equivalent decay structure such that $d\sim d_0$, there exist a family of isomorphisms $F_{d}:\mathcal{C}_d \rightarrow \mathcal{C}_{d_0}$ where
    \begin{equation}
        F_{d}(c)= \sum_{q=1}^m a_q\lambda_{d_0}(\widehat{c}_q)\widehat{c}_q
    \end{equation}
    where $c \in \mathcal{C}_d$ with basis elements $\widehat{c}_q$ and $\lambda_{d_0}(\widehat{c}_q)$ is a scaling of the basis element 
\end{lemma}
\begin{proof}
    Two coincidences algebras $C_d$ and $C_{d_0}$ are algebra isomorphisms if
    \begin{equation}
        F_{d}(\widehat{p}_i\cdot_d \widehat{p}_j) = F_{d}(\widehat{p}_i)\cdot_{d_0} F_{d}(\widehat{p}_j).
    \end{equation}
    which extends to the entire coincidence space. The LHS gives
    \begin{equation}
        F_{d}(\widehat{p}_i\cdot_d \widehat{p}_j)= \lambda_{d_0}(\widehat{p}_i\otimes \widehat{p}_j)c_{d}(\widehat{p}_i,\widehat{p}_j)\widehat{p}_i\otimes \widehat{p}_j,
    \end{equation}
    and the RHS gives
    \begin{equation}
        F_{d}(\widehat{p}_i)\cdot_{d_0} F_{d}(\widehat{p}_j) =\lambda_{d_0}(\widehat{p}_i)\lambda_{d_0}(\widehat{p}_j)c_{d_0}(\widehat{p}_i,\widehat{p}_j)\widehat{p}_i\otimes \widehat{p}_j. 
    \end{equation}
    The scalars in the isomorphism may be determined by following the graded structure of the coincidence space. Letting the scalars of coincidence degree one (transitions), $\lambda_{d_0}(\widehat{p}_i)$ and $\lambda_{d_0}$ be free, the scalars for higher degree coincidences may be determined.
    \begin{eqnarray}
        \lambda_{d_0}(\widehat{p}_i\otimes \widehat{p}_j) = \lambda_{d_0}(\widehat{p}_i)\lambda_{d_0}(\widehat{p}_j)\frac{c_{d_0}(\widehat{p}_i,\widehat{p}_j)}{c_{d}(\widehat{p}_i,\widehat{p}_j)}
    \end{eqnarray}
    Using Lemma~\ref{lemma:cocycle} and by setting the free transition scaling parameters to one, we may extend to higher degree coincidence:
    \begin{equation}
        \lambda_{d_)}(\widehat{p}_i\otimes \widehat{p}_j\otimes...\otimes\widehat{p}_{q}\otimes \widehat{p}_n) = \frac{c_{d_0}(\widehat{p}_i,\widehat{p}_j)...c_{d_0}(\widehat{p}_{q},\widehat{p}_n)}{c_{d}(\widehat{p}_i,\widehat{p}_j)...c_{d}(\widehat{p}_{q},\widehat{p}_n)}
    \end{equation}
    That is, for structurally equivalent coincidence algebras, since variation in the algebra is only induced via a change in the scalar connection that keeps it positive, an isomorphism can be established by canceling out this scaling from the scalar connection with a $\lambda$ determined for each coincidence element in $\mathcal{C}$. Therefore all structurally equivalent coincidence algebras are isomorphic and the isomorphism is continuous given $c_d$ is continuous and $c_d > 0$ for all $d \in B_{d_0}$.
\end{proof}
\end{definition}

\begin{theorem}
    The disjoint union of all structurally equivalent coincidence vector spaces $\mathcal{C}_d$ for $d \in B_{d_0}$ denoted,
    \begin{equation}
        \mathcal{A} = \bigsqcup_{d\in B_{d_0}}\mathcal{C}_d
    \end{equation}
    where $d_0$ is a fixed decay vector admits a trivial algebra bundle $\pi:\mathcal{A}\rightarrow B_d$ up to isomorphisms $\{F_d\}_{d\in B_{d_0}}$ such that
    \begin{equation}
        \mathcal{A} \cong B_{d_0}\times \mathcal{C}_{d_0}.
    \end{equation}
    with $\pi^{-1}(d) = \mathcal{C}_d$
\end{theorem}
\begin{proof}
    $\mathcal{A}$ is a trivial algebra bundle if a global trivialization $\Psi$ can be found such that 
    \begin{equation}
        \Psi : \mathcal{A}\rightarrow B_{d_0}\times \mathcal{C}_{d_0}.
    \end{equation}
    The family of  continuous isomorphisms $\{F_d\}_{d\in B_{d_0}}$ may be used to complete this mapping with
    \begin{equation}
        \Psi(c) = \left(\pi(c), F_{\pi(c)}(c)\right)
    \end{equation}
    for $c \in \mathcal{C}_d$. That is, a global trivialization can be found that is just the continuous rescaling of a fixed product bundle $B_{d_0}\times \mathcal{C}_{d_0}$ via the isomorphism $F_d$ for all $d \in B_{d_0}.$
\end{proof}
When considering a decays vectors coincidence probabilities that are calculated in its fiber, we require sections that lift that decay vector into its fiber in a manner that copies the vector into the space of the fiber -- which is always allowed given $k\mathcal{D} \subset\mathcal{C}$. We then denote the section $\tilde{s} :B_{d_0}\rightarrow\mathcal{A}$ such that $ d\mapsto (d,s(d))$. The underlying vector bundle of the algebra bundle is a product bundle $B_{d_0}\times\mathcal{C}$, thus elements of the bundle are tuples $(a,b)$ with $a\in B_{d_0}$ and $b\in\mathcal{C}$. However, when considering a vector in the bundle, the base space vector is what defines the algebra; what is being multiplied is the vector in $\mathcal{C}$, i.e., multiplying two vectors in a fiber defined at point $a\in B_{d_0}$ is defined as
\begin{equation}
   b\cdot_ac =(a,b)(a,c) = (a, b\cdot c) .
\end{equation}
With this is in mind, further notation restricts to writing the coincidence vector only with the path vector defined in the product. With the connection between levels and transitions now defined within the algebra, the probability vector of a given transition $\Gamma_1(d)$ and for two transitions in coincidence $\Gamma_2(d)$ can be given by
\begin{equation}\label{eqn:eqn41}
        \Gamma_1(d) = \tilde{b}\cdot_d \tilde{\tau},
\end{equation}
\begin{equation}\label{eqn:eqn42}
        \Gamma_{2}(d) = \tilde{b}\cdot_d \tilde{\tau} \cdot_d \tilde{\tau}.
\end{equation}
Where $\tilde{b} = (d,b(d))$---the branching section evaluated at $d$,  is the branching vector in the fiber at $d$ and $\tilde{\tau}=(d,\tau(d))$---the transition section evaluated at $d$ , is the transition vector in the same fiber, where $b(d)$ and $\tau(d)$ are defined in Equation 24. As with the path algebra, the order of multiplication is important and must be ordered in the hierarchy of the quiver, that is, the first term contains the branching ratios on the stationary paths which is connected to all paths below it via the scalar connection defined in the coincidence algebra. To get the coincidence probabilities we then further connect it to all paths below it; this is done by simply \textit{coincidence multiplying} onto the (lifted) transition vector again. What is different now from the path algebra formulation is that we now have terms in the probability vector that correspond to coincidences of paths that are not directly connected; e.g., with this a term for the coincidence of $x_{32}$ and $x_{10}$ is now given. Not only that, but for each transition, the level at which it decayed from is now a separate basis vector as the transition is now tensored with the vertices. Examples for $\Gamma_1(d)$ and $\Gamma_2(d)$ for a general decay scheme with $n=3$ is shown in Tables \ref{tab:tab1} and \ref{tab:tab2}, respectfully.
\begin{table}
     \centering
    \begin{tabular}{cccc}
    \hline
   Basis Vector& $f_3$& $f_2$& $f_1$\\
    \hline 
     $\widehat{p}_{30}$&$f_3x_{30}$ & &\\
     $\widehat{p}_{31}$&$f_3x_{31}$ & &\\
     $\widehat{p}_{32}$&$ f_3x_{32}$ & &\\
     $\widehat{\epsilon}_3\otimes\widehat{p}_{21}$&$f_3x_{32}x_{21}$ & &\\
     $\widehat{\epsilon}_3\otimes\widehat{p}_{20}$&$f_3x_{32}x_{20}$ & &\\
     $\widehat{\epsilon}_3\otimes\widehat{p}_{10}$&$f_3(x_{31}+x_{32}x_{21})x_{10}$ & &\\
     $\widehat{p}_{20}$&&$f_2x_{20}$ &\\
     $\widehat{\epsilon}_2\otimes\widehat{p}_{10}$&&$f_2x_{21}x_{10}$ &\\
     $\widehat{p}_{10}$&&&$f_1x_{10}$ \\
     \hline
    \end{tabular}
    \caption{Probability vector $\Gamma_1(d)$ coefficients for a 4 level decay quiver. The basis vectors determine the transition and the coefficients give the probability of that transition for a given decay. The basis vectors are separated by the branch they decayed from. }
    \label{tab:tab1}
\end{table}
\begin{table}
     \centering
    \begin{tabular}{ccc}
    \hline
   Basis Vector& $f_3$& $f_2$\\
    \hline 
     $\widehat{p_{31}}\otimes\widehat{p_{10}}$&$f_3x_{31}x_{10}$ & \\
     $\widehat{p_{32}}\otimes\widehat{p_{20}}$&$ f_3x_{32}x_{20}$ & \\
     $\widehat{p}_{32}\otimes \widehat{p}_{10}$&$ f_3x_{32}x_{21}x_{10}$ & \\
     $\widehat{p_{32}}\otimes\widehat{p_{21}}$&$ f_3x_{32}x_{21}$ & \\
     $\widehat{\epsilon}_3\otimes\widehat{p_{21}}\otimes\widehat{p_{10}}$&$f_3x_{32}x_{21}x_{10}$  &\\
     $\widehat{p_{21}}\otimes\widehat{p_{10}}$ &&$f_2x_{21}x_{10}$ \\
     \hline
    \end{tabular}
    \caption{Probability vector $\Gamma_2(d)$ coefficients for a 4 level decay quiver. The basis vectors determine the coincidence and the coefficients give the probability of that coincidence for a given decay. The basis vectors are separated by the branch they decayed from. }
    \label{tab:tab2}
\end{table}We also point out how easily gated probability vectors emerge out of the coincidence algebra--indeed, that is its strength. The transition matrix formalism allows for the summing corrections for \textit{all} of the  \textit{singles} spectra (single gamma-ray peaks) in the form of a matrix. As shown in equation 6, the coincidence probability for two \textit{specific} gamma can be given; however, there is no equation the produces all possible coincidences (such would require a rank for tensor, with two indices for each gamma). In \cite{schmidt2025measurabilitytruecoincidencesumming}, they showed how to calculate the final probability matrix with summing corrections for a given gate (gamma in coincidence) by partitioning the transition matrix into block matrices and treating the block matrices with the Semkow formalism separately before direct summing them back together. In the coincidence algebra, all possible coincidences can be given by single vector. To gate on a certain gamma, $\Gamma_C(d)$ can be projected onto a subspace defined by the gate. The partitioned Semkow formalism in \cite{schmidt2025measurabilitytruecoincidencesumming} indeed gives the transition matrix representation of $\Gamma_C(d)$ projected onto a gated subspace. Projections of coincidence vectors can be defined in the following manner. First, we begin by formulating how to compare tensor products of paths. This can be done by defining forms that sum over the paths in a tensor product:
\begin{equation}
    [\widehat{p}_i, \widehat{p}_1\otimes\widehat{p}_2\otimes...\otimes\widehat{p}_j]_p =\sum_{k=1}^j \langle\widehat{p}_k,\widehat{p_i} \rangle_p,
\end{equation}
which expands linearly for a given vector $c\in\mathcal{C}$ by
\begin{equation}
    [\widehat{p}_i, c]_p = \sum_{q=1}^m a_q[\widehat{p}_i,\widehat{c}_q]_p,
\end{equation}
where
\begin{equation}
    c = \sum_{q=1}^ma_q\widehat{c}_q,
\end{equation}
with $\widehat{c}_q$ denoting some tensor product of paths (a basis vector in $\mathcal{C}$). We therefore define the following gate projector
\begin{definition}
    The gate projector $G_1(\widehat{p}_i, \widehat{c}):\mathcal{C}\rightarrow\mathcal{C}$ projects a vector $c\in\mathcal{C}$ onto the subspace whose basis vectors overlap with a path $\widehat{p}_i$ such that $[\widehat{p}_i,\widehat{c}]_p> 0$.
    \begin{equation}
        G_1(\widehat{p}_i, \widehat{c}) = \sum_{q=0}^m a_q[\widehat{p}_i,\widehat{c}_q]_p\widehat{c}_q
    \end{equation}
\end{definition}For example, given the coincidence vector in Table \ref{tab:tab2}, 
\begin{align}
G_1(\widehat{p}_{32}, \Gamma_2(d)) =&     f_3x_{32}x_{20}\widehat{p_{32}}\otimes\widehat{p_{20}} + f_3x_{32}x_{21}x_{10}\widehat{p}_{32}\otimes \widehat{p}_{10}\\\nonumber
&+f_3x_{32}x_{21}\widehat{p_{32}}\otimes\widehat{p_{21}}
\end{align}
The gate projector may be extended to gate on coincidences, not just single transitions:
\begin{equation}
    G_2(\widehat{p}_i, \widehat{p}_l, \widehat{c})=\sum_{q=0}^m a_q[\widehat{p}_i,\widehat{c}_q]_p[\widehat{p}_l,\widehat{c}_q]_p\widehat{c}_q
\end{equation}
This can of course easily be extended to higher order coincidences by multiply on a form for each given transition.
Let us now consider how detection maps work in the base space on the coincidence algebra bundle. 

\subsection{Detection maps in the bundle.}
With the detection maps taking a decay vector to a new vector in the path algebra that alters the probabilities to account for detection effects, when we calculate coincidence probabilities in that detection mapped fiber, the probabilities now alter the coincidence multiplication. It is sufficient for now to realize that we may transport a vector in one fiber to another fiber with out having to consider a connection given that a trivial connection can be established on the underlying product vector bundle\footnote{A multiplicative connection would allow for the transportation of vectors in a way that scales them to cancel out the change in the algebra. This could be done with the isomorphisms we defined in the global trivialization of the coincidence algebra bundle. However, we require the scaling here as it defines the correct probabilities.}. This is important when dealing with multiple detection maps to calculate a given probability vector. Whenever we are multiplying paths together, it must be defined within the fiber of a given point in the base space, i.e, it must be defined for a given decay vector. When taking the coincidence product in a detection mapped fiber, the probabilities that connect the two paths in the product will be the ones defined by the detection map. For transitions of interest, a detection map that accounts for full energy peak efficiencies would be desired. However, the probabilities that connect the transitions of interest would be from a detection mapped decay vector that accounts for summing out probabilities. Therefore, we wish to calculate summing out probabilities in the fiber at the summing out mapped decay vector, but use vectors transported from other fibers in the base space (the full energy ones). We may therefore have maps between points in the base space, such as $d\mapsto\varphi(d)$, and maps between fibers, $(\varphi(d),\varphi(d))\mapsto (\varphi^\prime(d), \varphi(d))$. Again, given the trivial product vector bundle underlying the algebra bundle, the manner in which we move vectors around does not matter, it is only the transportation of products that changes. Put more aptly, the detection maps are not homomorphisms of the coincidence algebra defined on the fibers. When considering the power expansion of detection mapped vectors calculated in the coincidence algebra, we must now distinguish what map is being applied to the vector being multiplied and what detection map is defining the coincidence multiplication. For lifting sections that have been mapped to detection vectors, we denote
\begin{equation}
    \tilde{\varphi}(s) := \left(\varphi(d),s(\varphi(d)\right).
\end{equation}
When considering the same problem, of coincidence summing with summing out and summing in, we may use the following power expansion where the fiber is now explicitly stated:

\begin{align}
       \overline{\tilde{\varphi}_h(\tau)}_N ^{n_{\varphi_o(\tau)}} = \sum_{k=1}^n \frac{\tilde{\varphi}_h(\tau) ^{k_{\varphi_o(\tau)}}}{N^{k-1}} =  \tilde{\varphi}_h(\tau) + \frac{\tilde{\varphi}_h(\tau)\cdot_{\varphi_o(\tau)}\tilde{\varphi}_h(\tau)}{N}+\frac{\tilde{\varphi}_h(\tau)\cdot_{\varphi_o(\tau)}\tilde{\varphi}_h(\tau)\cdot_{\varphi_o(\tau)}\tilde{\varphi}_h(\tau)}{N^2} +...
\end{align}

Note, we begin the power expansion at $k=1$ as the coincidence algebra does not have a global identity, only a complete set of idempotents. Regardless though, for the power expansion defining the paths of interest in the probability vector, there should be no $k=0$ term as we only want path terms. Conversely, in the scalar connection -- which is calculated in the path algebra and thus does have a global identity available -- the power expansion must start at $k = 0$ to allow for trivial scalar connections between paths that can concatenate\footnote{Although, as shown in equation 31, summing-in expansion start at $k=1$}. It is also needed for assigning the proper branching ratios. With the detection maps established in the coincidence algebra bundle we may now calculate the probability vector with summing in and summing out maps:

\begin{equation}\label{eqn:eqn50}
    \Gamma_1(d) = \tilde{b}\cdot_{\varphi_o(\tau)}\overline{\tilde{\varphi}_h(\tau)} ^{n_{\varphi_o(\tau)}} \cdot_{\varphi_o(\tau)}\widehat{\epsilon_0}.
\end{equation}

The transitions of interest are given in the center term where summing in is accounted for with potential summing out between disconnected paths. Each path is connected to their possible branching levels above and to the ground state below through the summing out probabilities connecting between.
As an example, the probability vector with detection maps for the 4 level decay quiver given by Eq. \ref{eqn:eqn50} is shown in Table 1. The basis vectors are separated by the number of full energy gammas they represent. The coefficients are shown in the columns and are distinguished by the branch they decayed from. 


\begin{table*}
     \centering
    \begin{tabular}{||c||c|c|c||}
    \hline
   Basis Vector& $f_3$& $f_2$& $f_1$\\
    \hline \hline
     $\widehat{\epsilon}_3\otimes\widehat{p}_{30}$&$f_3h_{30}$ & &\\
     $\widehat{\epsilon}_3\otimes\widehat{p}_{31}\otimes\widehat{\epsilon}_0$&$f_3h_{31}o_{10}$ & &\\
     $\widehat{\epsilon}_3\otimes\widehat{p}_{32}\otimes\widehat{\epsilon}_0$&$ f_3h_{32}o_{20}$ & &\\
     $\widehat{\epsilon}_3\otimes\widehat{p}_{21}\otimes\widehat{\epsilon}_0$&$f_3o_{32}h_{21}o_{10}$ & &\\
     $\widehat{\epsilon}_3\otimes\widehat{p}_{20}$&$f_3o_{32}h_{20}$ & &\\
     $\widehat{\epsilon}_3\otimes\widehat{p}_{10}$&$f_3(o_{31}+o_{32}o_{21})h_{10}$ & &\\
     $\widehat{\epsilon}_2\otimes\widehat{p}_{20}$&&$f_2h_{20}$ &\\
     $\widehat{\epsilon}_2\otimes\widehat{p}_{10}$&&$o_{21}h_{10}$ &\\
     $\widehat{\epsilon}_1\otimes\widehat{p}_{10}$&&&$f_1h_{10}$ \\
     \hline
     $\widehat{\epsilon}_3\otimes\widehat{p_{31}}\otimes\widehat{p_{10}}$& $f_3h_{31}h_{10}/N$& &\\
     $\widehat{\epsilon}_3\otimes\widehat{p_{32}}\otimes\widehat{p_{20}}$& $f_3h_{32}h_{20}/N$& &\\
     $\widehat{\epsilon}_3\otimes\widehat{p}_{32}\otimes\widehat{p}_{10}$&$f_3 h_{32}o_{21}h_{10}/N$ & &\\
     $\widehat{\epsilon}_3\otimes\widehat{p_{32}}\otimes\widehat{p_{21}}\otimes\widehat{\epsilon}_{0}$&$f_3h_{32}h_{21}o_{10}/N$& &\\
     $\widehat{\epsilon}_3\otimes\widehat{p_{21}}\otimes\widehat{p_{10}}$&$f_3o_{32}h_{21}h_{10}/N$& &\\
     $\widehat{\epsilon}_2\otimes\widehat{p_{21}}\otimes\widehat{p_{10}}$&&$f_2h_{21}h_{10}/N$ &\\
     \hline
     $\widehat{\epsilon}_3\otimes\widehat{p_{32}}\otimes\widehat{p_{21}}\otimes\widehat{p_{10}}$&$f_3h_{32}h_{21}h_{10}/N^2$ & &\\
     \hline
    \end{tabular}
    \caption{Probability vector $\Gamma(d)$ coefficients for a 4 level decay quiver with summing out probabilities $o_{ij} = \varphi_o(x_{ij})$ and detected transition $h_{ij} = \varphi_h(x_{ij})$. The basis vectors are separated by the branch they decayed from. The first block contains the basis vectors where 1 gamma is detected at full energy; the second block contains the basis vectors where 2 gammas are detected at full energy; the third block contains the only basis vector where 3 gammas are detected at full energy. }
    \label{tab:placeholder}
\end{table*}
To calculate the detected coincidence probability vector $\Gamma_2(d)$ with summing corrections, we must first define a new detection map that allows for the second gamma (with the first being the gate) to be detected--that is, it does not sum with the first gamma. To do this we continue to use the isotropic approximation and say the probability that the second gamma does not sum with the first is given by $(N-1)/N$. We therefore define the gated hit detection map as
\begin{equation}
    \varphi_g = \frac{\varphi_h (N-1)}{N}.
\end{equation}
Again, with $N$ being the number of detectors. The detected coincidence probability vectors is therefore
\begin{equation}\label{eqn:eqn52}
     \Gamma_2(d) = \tilde{b}\cdot\overline{\tilde{\varphi}_h(\tau)} ^{n_{\varphi_o(\tau)}}\cdot\overline{\tilde{\varphi}_g(\tau)} ^{n_{\varphi_o(\tau)}} \cdot\widehat{\epsilon_0},
\end{equation}
where the multiplication $\cdot :=\cdot_{\varphi_o(\tau)}$ is over the summing out fiber as is done in Eq.\ref{eqn:eqn50}. However, with higher order coincidence that account for the summing in there is a degeneracy in the basis vectors for different detection scenarios. For example, the basis vector $\widehat{p_{32}}\otimes\widehat{p_{21}}\otimes\widehat{p_{10}}$ could correspond to the summing-in to $\widehat{p}_{31}$ in coincidence with $\widehat{p}_{10}$,  or $\widehat{p}_{32}$ in coincidence with the summing-in to $\widehat{p}_{20}$. Calculating the coincidence vector as shown in Eq.\ref{eqn:eqn52} would provide the probability for both of these distinct detection scenarios onto the same basis vector. To avoid this detection degeneracy ( distinct detection events possessing the same basis vector), a gate projector can be applied to the gated hit detection map vector after power expansion. Before moving onto validation of the algebra with simulation data, another detection map needs to be established. To accurately compare to simulation data internal conversion must be accounted for. In such a case, the gamma is not emitted from the level and transition occurs by providing the energy to electrons. Internal conversion maps can easily be applied by dividing the transition probability for $1+\alpha$, with $\alpha$ being the internal conversion coefficient. Such a map may be defined as
\begin{equation}
    \varphi_{IC}(\tau) = \sum_{\widehat{p} \in \mathcal{P}_1}\frac{1}{1+\alpha_{\widehat{p}}}\langle\widehat{p},\tau \rangle_p \widehat{p},
\end{equation}
With considering internal conversion, one must distinguish between transition probabilities and emission probabilities. The former defines the probability that the nucleus simply transitions to the lower level, regardless of a gamma being emitted. The latter requires gamma emission and thus must have internal conversion coefficients mapped into the decay vector. 
\section{Validation with Geant4 Simulations}

The decay quiver formalism to calculate emission and detection probabilities using the path algebra and the coincidence algebra for nuclear level schemes has been implemented into the \texttt{COINAlgebra} C++ Package.\footnote{A beta version is released on GitHub, the coding package is still in its early stages and is in ongoing development at the time of this paper.} The package uses the formalism to define classes such as \texttt{DecayLevel}, \texttt{DecayTransition}, \texttt{DecayQuiver}, \texttt{Decay Probability}, \texttt{PathAlgebra}, and \texttt{CAlgebra} to calculate probabilities for a constructed decay quiver. The decay quiver formalism is then validated by comparison to Geant4 simulated gamma spectra. A Geant4 simulation was done that uses the detector geometry and properties of the GRIFFIN spectrometer \cite{Garnsworthy_2019}. 

The GRIFFIN spectrometer is a multi-detector array of HPGe clover detectors. Each of the 16 clovers composing the entire array is made of 4 n-type HPGe crystals, which makes up a square face of the 18 sides of a rhombicuboctahedron; the remaining two squares being used for the in-line radioactive beam which deposits nuclei onto a moving tape system that removes long-lived activity from the chamber at the end of each measurement cycle \cite{Garnsworthy_2019}. A plot of its simulated peak efficiency is shown in Fig.~\ref{fig:fig5} where a total efficiency of 0.7 is used for all energies (equating to the solid angle of the detector).\footnote{11 cm, Single crystals, no shields, US+DS SCEPTAR, no 
delrin}. All calculations with detection maps dependent on efficiencies in this section use the efficiencies provide by the GRIFFIN efficiency calculator \cite{Mills2015}, which is created using the same simulation setup \texttt{COINAlgebra} is benchmarked to. 

\begin{figure}[H]
    \centering
    \includegraphics[width=0.7\linewidth]{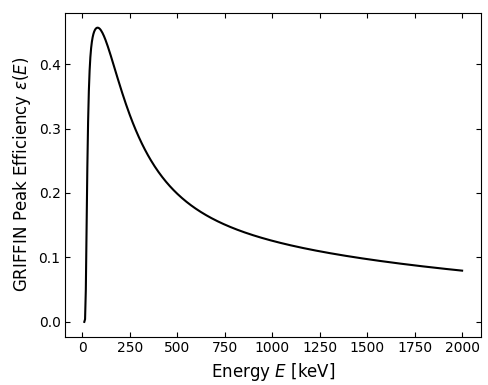}
    \caption{Simulated peak efficiencies for GRIFFIN \cite{Mills2015}.}
    \label{fig:fig2}
\end{figure}

A Geant4 simulation of the electron capture of $^{133}$Ba was done to validate the predictions of the decay quiver formalism with $10^7$ parent decays. The decay scheme data used for the simulation, as shown in Figure \ref{fig:fig5} and the \texttt{COINAlgebra} calculations comes from ENDSF shown in Table \ref{tab:ba133_decay_scheme}\cite{ENSDF_Ba133}. Geant4 simulations serve as a good validation for the decay quiver formalism because the simulations include true coincidence summing affects and detector efficiency as the only corruption of emission data which are phenomena are included in the formalism. From the output of the simulation, single $\gamma$ spectra and $\gamma-\gamma$ coincidence spectra were created to compare to the \texttt{COINAlgebra} predictions. For the singles spectra, the path algebra is sufficient in predicting the probabilities. The first comparison constructed is the output of Equation 32, which computes the detection probabilities for single $\gamma$-rays with summing-in and summing-out effects included. The functions and formalism contained in that equation were implement through the \texttt{DecayPath}, \texttt{DecayVector}, \texttt{DecayProbability}, and \texttt{DetectionMap} classes which provide the detection probabilities account for feeding from above with summing-in and summing-out. The simulation was done to create spectra with perfect resolution, such that detection counts are contained in single bins of the spectra where background subtraction can be applied by subtracting off the average of the neighboring bins (ensuring no true $\gamma$ peaks are neighboring which is easily done with a high bin count of 0.4 keV).

\begin{figure}[]
    \centering
    \includegraphics[width=0.7\linewidth]{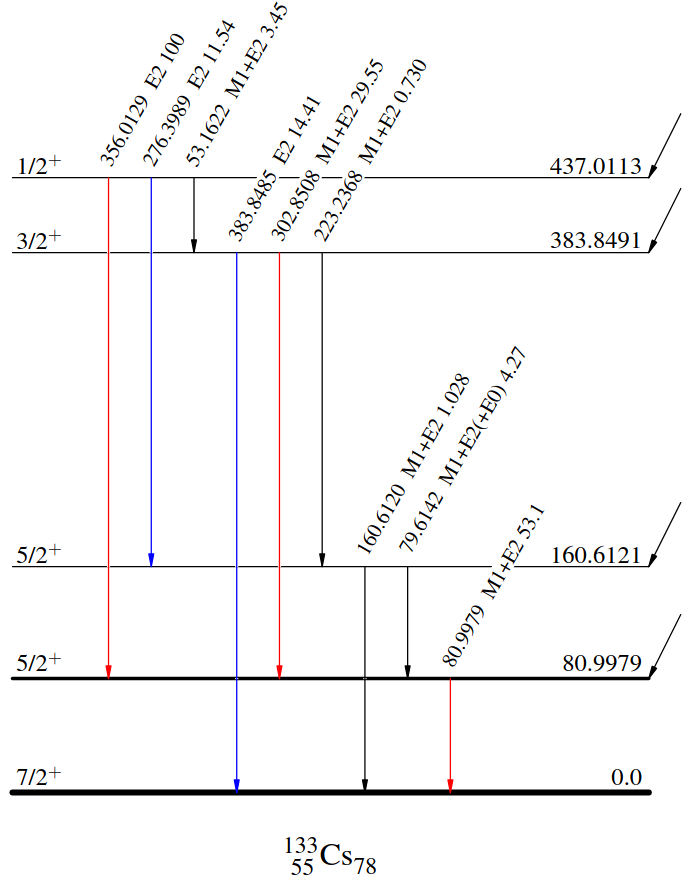}
    \caption{Gamma decay levels scheme from the electron capture of $^{133}$Ba. Figure from ENSDF \cite{ENSDF_Ba133}.Initial populations of the $^{133}$Cs levels following electron capture decay of $^{133}$Ba used in the \texttt{COINAlgebra} calculation are 0.855 for branching to level 437 keV and 0.145 for branching to level 383 keV. The lower levels have branching limits $<0.002$ and $<0.001$ and thus have been set to zero for the calculation.}
    \label{fig:fig5}
\end{figure}
\begin{figure}[]
    \centering
    \includegraphics[width=0.7\linewidth]{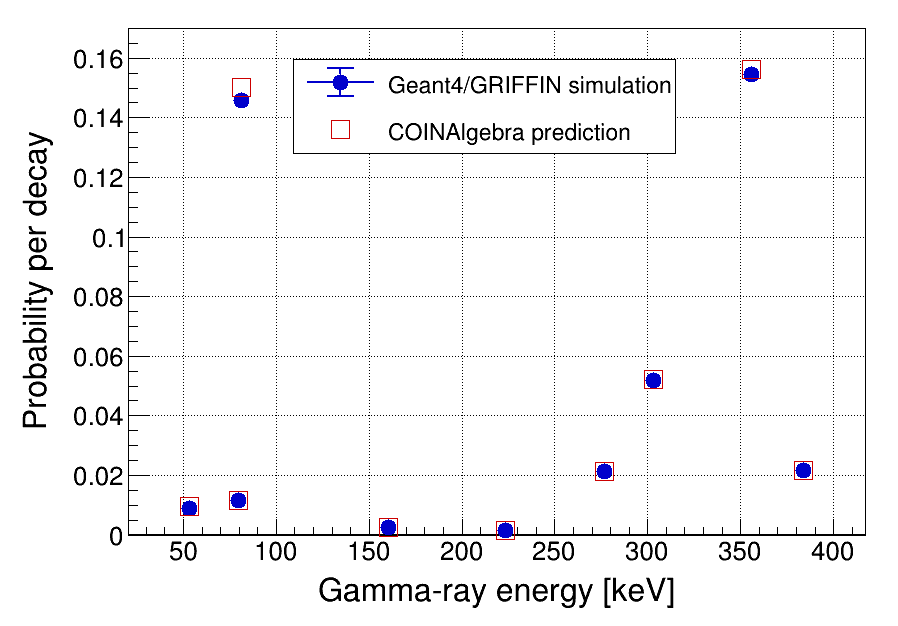}
    \label{fig:fig7}
    \caption{Comparison of the simulated and COINAlgebra-predicted full-energy detection probabilities for selected $\gamma$-ray transitions following the decay of $^{133}$Ba.}
\end{figure}
\begin{figure}[]
    \centering
    \includegraphics[width=0.7\linewidth]{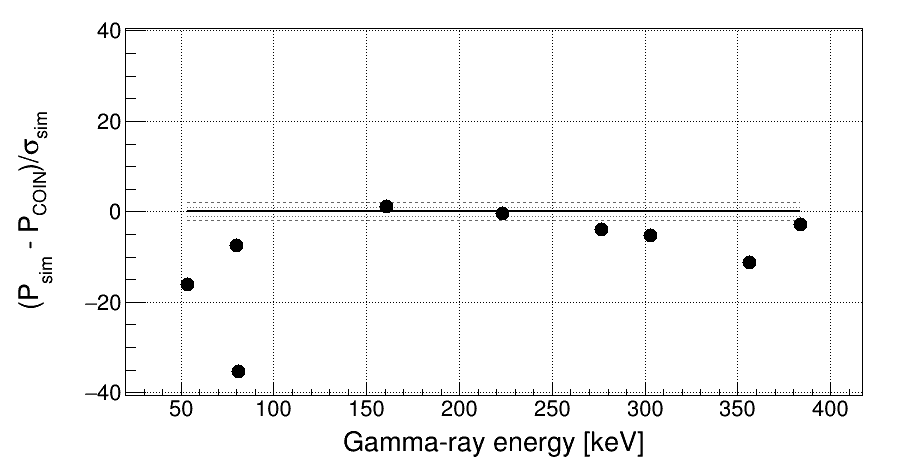}
    \caption{Pulls between the Geant4/GRIFFIN simulated and COINAlgebra-predicted full-energy singles detection probabilities for selected $\gamma$-ray transitions following the decay of $^{133}$Ba.}
    \label{fig:fig8}
\end{figure}

 The results of the \texttt{COINAlgebra} prediction along with the results from the Geant4 simulation in GRIFFIN for the single $\gamma$ detection probabilties are shown in Figure \ref{fig:fig7}. The pulls between the two datasets, being the residuals weighted by the uncertainty given to the simulation data ( set to $\sqrt{N}$ where $N$ is the bin count) is shown in Figure \ref{fig:fig8}. As an initial benchmark, the prediction of the formalism are in general agreement with the detection probabilities seen in the simulated data. We recognize the relatively large residuals in comparison to the uncertainty of the simulated data, though any discrepancy between the results we agree must come from possible differences in the efficiencies used in the calculations versus those present in the particular simulation run. This validation is meant to serve merely as a benchmark for the algebra defined herein. A possible source of discrepancy may be coming from the isotropic approximation used to calculate the summing-in and summing-out probabilities. This is apparent in the large discrepancy of the 81 keV as it is a strong transition that will encounter a greater degree of summing out from the $\gamma$-rays above. 

\begin{table}[]
    \centering
    \label{tab:ba133_decay_scheme}
    \begin{tabular}{c c c c c c}
        \hline
        $E_i$ [keV] &
        $E_f$ [keV] &
        $E_{\gamma}$ [keV] &
        $P_{\gamma+\mathrm{IC}}$ &
        $\alpha$ &
        $P_{\gamma}$ \\
        \hline
         81.00 &   0.00 &  81.00 & 1.00 & 1.70 & 0.37 \\
        160.61 &  81.00 &  79.61 & 0.90 & 1.77 & 0.32 \\
        160.61 &   0.00 & 160.61 & 0.10 & 0.29 & 0.08 \\
        383.85 & 160.61 & 223.24 & 0.02 & 0.10 & 0.02 \\
        383.85 &  81.00 & 302.85 & 0.67 & 0.04 & 0.64 \\
        383.85 &   0.00 & 383.85 & 0.32 & 0.02 & 0.31 \\
        437.01 & 383.85 &  53.16 & 0.17 & 5.66 & 0.03 \\
        437.01 & 160.61 & 276.40 & 0.09 & 0.06 & 0.08 \\
        437.01 &  81.00 & 356.01 & 0.74 & 0.03 & 0.72 \\
        \hline
    \end{tabular}
    \caption{Electromagnetic transitions in $^{133}$Cs used in the
    \texttt{COINAlgebra} calculation following the electron-capture decay of
    $^{133}$Ba. The transition probability $P_{\gamma+\mathrm{IC}}$
    represents the total electromagnetic branch, including both
    $\gamma$ emission and internal conversion. The internal conversion
    coefficient is denoted by $\alpha$, and the corresponding
    $\gamma$-emission probability is calculated as
    $P_{\gamma}=P_{\gamma+\mathrm{IC}}/(1+\alpha)$.}
\end{table}

The coincidence algebra is implemented into the \texttt{DecayCoin} and \texttt{CAlgebra} classes where detection probabilities can be calculated using the same detection maps defined for the path algebra. The coincidence algebra is particularly powerful because it can reproduce the detected spectrum of gamma-rays by including the summing of all possible gammas--the path algebra, and thus transition matrices only create calculate full energy summing for composable transitions. The probability vector produced by Equation 70 gives all the single transition peaks with summing-out accounted for, the composable summing-in peaks which would corrupt the single transition peaks (but are kept separate by the higher degree basis vectors), and the summed peaks from disconnected transitions. For this reason, Equation 70 is seen as the \textit{spectrum producing probability vector}. This is precisely why the path algebra requires extension to the coincidence algebra. 

\begin{table}[]
    \centering
    \label{tab:noncomposable_sum_peaks}
    \begin{tabular}{ccc}
        \hline
        \hline
        Sum peak (keV) &
        $E_{\gamma_1}$ (keV) &
        $E_{\gamma_2}$ (keV) ) \\
        \hline
        132.77 & 53.16  & 79.61 \\
        134.16 & 53.16  & 80.99 \\
        213.77 & 53.16  & 160.61 \\
        304.23 & 223.23 & 80.99 \\
        357.39 & 276.39 & 80.99 \\
        \hline
        \hline
    \end{tabular}
    \caption{Non-composable full-energy summing peaks predicted for the
    $^{133}$Ba decay scheme. The listed input energies correspond to detected
    $\gamma$ rays whose coincidence factors are not directly composable in the
    ordinary path algebra.}
\end{table}
This is useful for large level schemes where many non composable transitions create full energy summing peaks, some of which by chance often create near doublets corrupting important peaks of interest which can easily go unnoticed. The decay of $^{133}$ Ba gives a clear example of this with the summing of the 79 keV and the 276 keV creating a near doublet corrupting the 356 keV peak. For high precision measurements in gamma-spectroscopy, these can contribute to systematic uncertainty in peak area evaluation if the near doublet corruption is significant enough. This is particularly the case in the $\beta^+$ decay of $^{22}$Mg where the 511 keV from positron annihilation sums with the 74 keV corrupting the 583 keV peak. This is discussed more in section 5 on auxiliary radiation.
\begin{figure}
    \centering
    \includegraphics[width=0.7\linewidth]{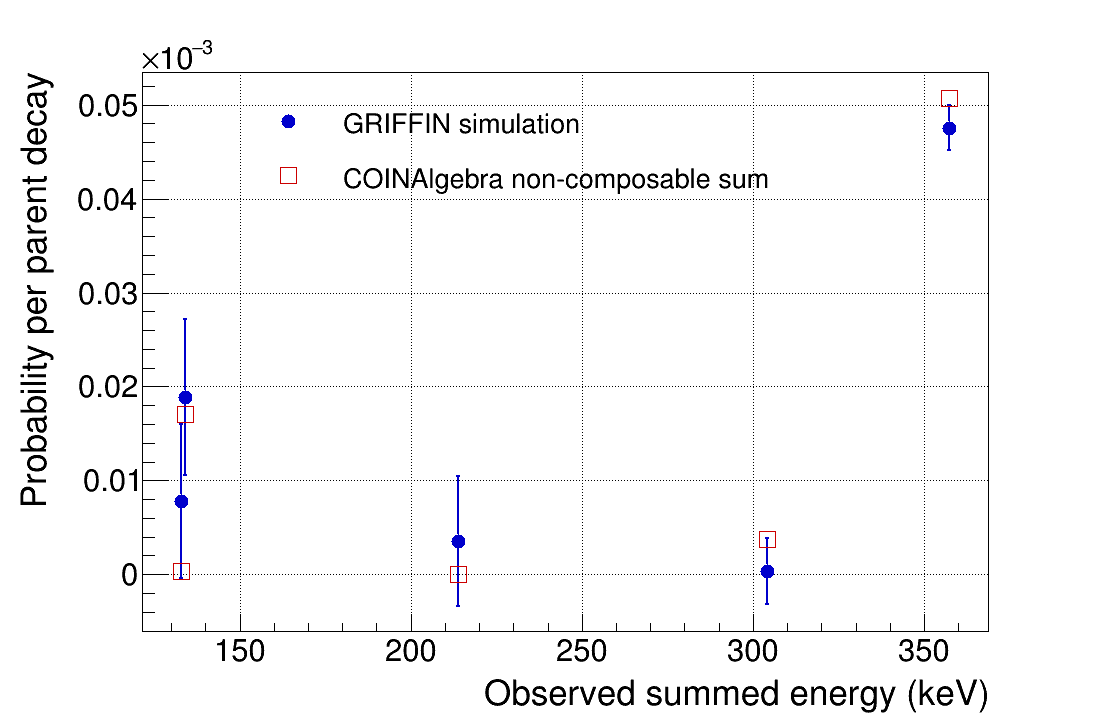}
    \caption{Comparison of the simulated and COINAlgebra-predicted full energy non composable summing peaks following the decay of $^{133}$Ba.}
    \label{fig:placeholder}
\end{figure}
\par 
The coincidence algebra, naturally, easily calculates coincidences probability for emission and detection. The detection probabilities of $\gamma-\gamma$ coincidences for all observable pairs of $\gamma$-rays are calculated and compared to the simulation coincidence probabilities. The results of this are shown in Figure 8, with the pulls showing any discrepancies between the predicted and simulated results. As previously stated, to include summing-in for transitions of interest in the coincidence probability vector $\Gamma_2(d)$ creates a degeneracy in the higher order paths. This removes the ability to calculate coincidence detection probabilities with summing-in \textit{with a total vector}. To calculate summing in for coincidences, we may simply select a gate on the coincidence and get the detected coincidence spectra for that gate. However, for the purpose of providing a benchmark for the algebra structure itself, the predictions in Figure 8 are done without summing-in expansions, and thus calculates
\begin{equation}
     \Gamma_2(d) =\left(\frac{N-1}{N} \right) \tilde{b}\cdot\overline{\tilde{\tau}} ^{n_{\varphi_o(\tau)}}\cdot\overline{\tilde{\tau}} ^{n_{\varphi_o(\tau)}} \cdot\widehat{\epsilon_0},
\end{equation}
Again, these validations are merely to confirm the algebra structure works and are not concerned with exactly reproducing the simulated values with high accuracy. The results in Figure 8 show good agreement with the simulated values even without summing-in corrections.
\begin{figure}[]
    \centering
    \includegraphics[width=0.7\linewidth]{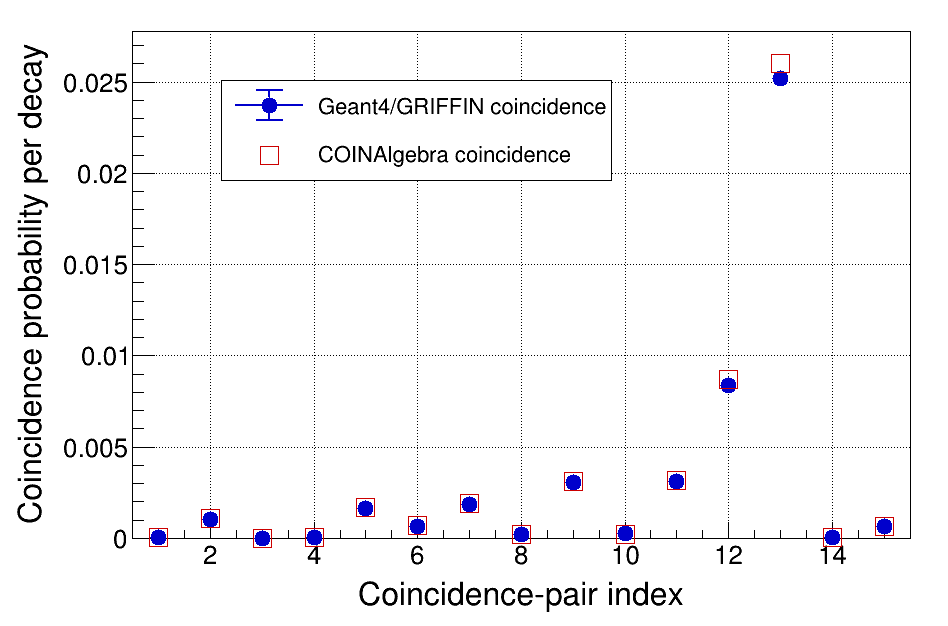}
    \caption{Comparison of the simulated and COINAlgebra-predicted full-energy coincidence probabilities for selected pairs of $\gamma$ rays following the decay of $^{133}$Ba. The coincidence-pair index refers to the energy combinations listed in Table~\ref{tab:ba133_coincidence_indices}.}
    \label{fig:placeholder}
\end{figure}

\begin{figure}[]
    \centering
    \includegraphics[width=0.7\linewidth]{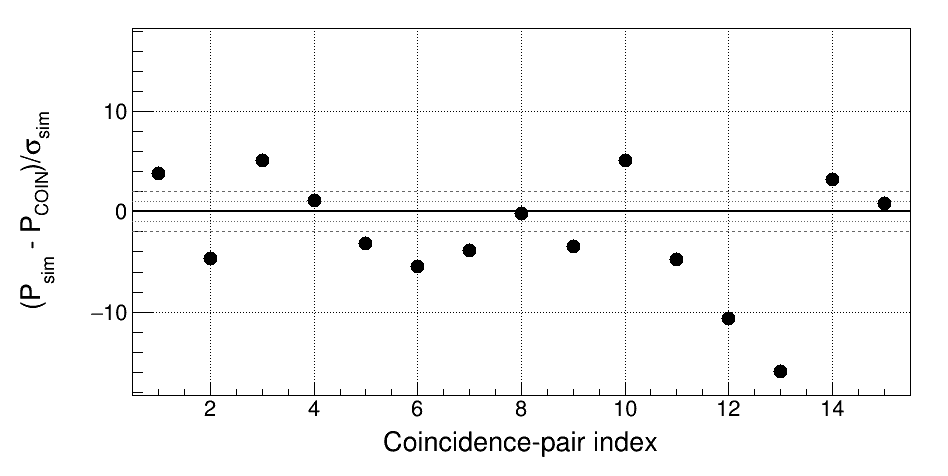}
    \caption{Pulls between the Geant4/GRIFFIN simulated and COINAlgebra-predicted two-fold full-energy coincidence probabilities for the selected $^{133}$Ba $\gamma$-ray pairs. The coincidence-pair index corresponds to the energy combinations listed in Table~\ref{tab:ba133_coincidence_indices}.}
    \label{fig:placeholder}
\end{figure}
We refrain from providing higher order coincidence comparison as it would not validate the algebra anymore. It is also the case that going higher order in coincidences events in the simulation decreases in orders of magnitude the event probability and the statistics on such rare events would not provide much meaningful data or validation. 
\begin{table}[]
    \centering
    \label{tab:ba133_coincidence_indices}
    \begin{tabular}{c c c}
        \hline
        Coincidence Index & $E_{\gamma,1}$ [keV] & $E_{\gamma,2}$ [keV] \\
        \hline
         1  & 53.16  & 79.61  \\
         2  & 53.16  & 80.99  \\
         3  & 53.16  & 160.61 \\
         4  & 53.16  & 223.23 \\
         5  & 53.16  & 302.85 \\
         6  & 53.16  & 383.84 \\
         7  & 79.61  & 80.99 \\
         8  & 79.61  & 223.23 \\
         9  & 79.61  & 276.39 \\
        10  & 80.99  & 223.23 \\
        11  & 80.99  & 276.39 \\
        12  & 80.99  & 302.85 \\
        13  & 80.99  & 356.01 \\
        14  & 160.61 & 223.23 \\
        15  & 160.61 & 276.39 \\
        \hline
    \end{tabular}
    \caption{Gamma-ray energies corresponding to each coincidence index used in the
    $^{133}$Ba coincidence comparison.}
\end{table}

\section{Remarks}
\subsection{Auxiliary Radiation}
A note should be said here on what motivated this framework. As previously mentioned, the transition matrix formalism for summing corrections in nuclear level schemes was initially put together by Semkow. Afterwards, others like \cite{KORUN1993478} extended it by including \textit{auxiliary radiation}; that is, others forms of radiation like X-rays from internal conversion and 511 keV gamma-rays from position-electron annihilation. It is typical that auxiliary radiation can be added on into the formalism by the use of virtual branches. In this case, branches are added on to the decay level scheme with the additional transitions representing the auxiliary radiation. There are multiple ways to do this corresponding to different types of radiation. For X-rays, \cite{KORUN1993478} showed that virtual branches can be placed between the levels to allow for internal conversion and X-rays. In \cite{MCCALLUM1975189}---predating the work of \cite{SEMKOW1990437}, 511 keV gammas are proposed to be incorporated as pseudo-branches above the level scheme, where internal conversion coefficients of $0.5$ double the probability of the emission allowing for the two 511 keV produced in annihilation--though not to be detected simultaneously given a single detector setup. For modern day gamma spectroscopy experiments, like those done with \cite{Garnsworthy_2019}, a spherical distribution of multiple detectors is more standard and allows for coincidence measurements. For multiple detector arrangements having two 511 keV gammas below the ground state, as shown in Figure 3, is more suitable. In this manner, we ensure that all decays give the 511 keV radiation from the positron annihilation. The transition matrix formalism can handle the summing out corrections with the 511 virtual branches but is restricted to only allow summing in corrections with the transitions directly connected above the ground state. It is this very need to allow for full energy summing in corrections with 511 keV gammas in \textit{general} that has motivated this work. However, proposed here is not just a framework that handles that particular case, but handles the combinatorics of all possible coincidences. An example where the 511 keV summing is important is the $\beta^+$ decay of $^{22}Mg$ where summing with 511 keV gammas corrupts one of the large peaks of interest \footnote{The precision measurement of the superallowed $\beta^+$ decay of $^{22}Mg$--a measurement that contributes to the $V_{ud}$ value in the CKM matrix--requires a precise measurement of this corrupted peak.}. However, the coincidence algebra not only addresses this by allowing for full energy summing-in between all gammas, but further separates coincidence terms by their branches and makes the extension to gated probabilities trivial.

\begin{figure}[]
    \centering
    \includegraphics[width=0.7\linewidth]{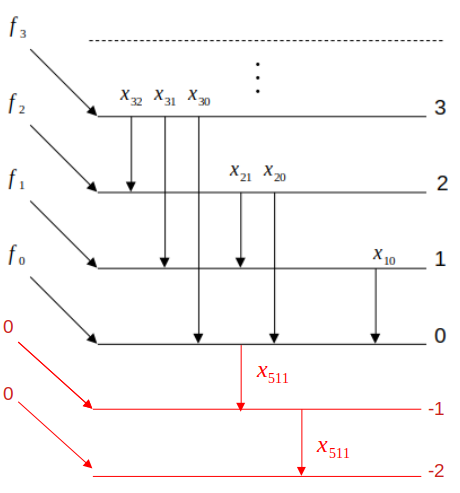}
    \caption{A general nuclear level scheme with virtual branches below the ground state to allow for 511 keV gammas in the summing formalism.}
    \label{fig:placeholder}
\end{figure}
\subsection{Uncertainty Propagation in the Path Algebra}
There is another aspect of the path algebra and the coincidence algebra that are particularly useful and provide a natural framework for understanding uncertainty propagation in the determination of feeding fractions, i.e., probabilities of transition for a given parent decay. Recent work by \cite{athanasios} uses similar graphical treatments of nuclear levels with the standard transition matrix treatment. They differ slightly be considering long lived states as absorbing states that don't terminate to the ground level. They also provide a good treatment of uncertainty propagation for the transition probabilities in the level scheme and how they contribute to the total variance for feeding probabilities. Such a treatment fits naturally into the path algebra, and can maybe even be extended further. As stated, the path algebra is isomorphic to the transition matrix algebra modded by paths of equivalent sources and targets. The separation of the different paths in the path algebra may allow for a further distinction between histories in the decay scheme and how they contribute differently to the total feeding fraction variance. In \cite{athanasios}, they provide fractional contributions to the total variance of a feeding probability from the different transitions above. In doing so, they are able to provide rankings on the transitions specific contributions to the total uncertainty of the feeding fraction. In the path algebra, with cascades separated into different basis vectors, the feeding fraction variance may also be decomposed into different cascade/decay histories. Moreover, the feeding fraction derivatives have a natural geometric interpretation in the path algebra vector space as elements of a covector. The probability vectors,
\begin{equation}
    \Gamma_P = V_t(b\cdot \bar{\tau}^n)\cdot\tau
\end{equation}
are vector fields over the path algebra vector space, with $P(\widehat{p}_i)$ as scalar fields. The vector contained in the target vertex projector separates feeding histories which are summed back together by the projection. For a decay vector,
\begin{equation}
    d = \sum_k x_k\widehat{p}_k
\end{equation}
The differential,
\begin{equation}
    dP(\widehat{p}_i) = \sum_j\frac{\partial P(\widehat{p}_i)}{\partial x_j}dx_j
\end{equation}
lives in the cotangent space of $k\mathcal{D}$. The variance of the probabilities  $P(\widehat{p}_i)$ with zero covariance in the transition probabilities is given by
\begin{equation}
    \sigma^2\left(P(\widehat{p}_i)\right) = \sum_j \frac{\partial P(\widehat{p}_i)}{\partial x_j}\sigma^2_j
\end{equation}
Thus, the sensitivity of the feeding fractions to uncertainties in the transition probabilities are represented within covectors in $k\mathcal{D}$. The different histories for the feeding from above a level $i$ are explicitly given through separate basis vectors in $T_i(b\cdot \bar{\tau}^n)$, where $T_i$ is the target projector. Thus the feeding $f_i$ to level $i$ is given by
\begin{equation}
    f_i = ||T_i(b\cdot \bar{\tau}^n)||,
\end{equation}
where $||\cdot||$ is the $l_i$ norm (sum of components). If we define $F_i=T_i(b\cdot \bar{\tau}^n)$ as the feeding vector for level i, written
\begin{equation}
    F_i = \sum_j a_j \widehat{p}_j
\end{equation}
with $t(\widehat{p}_j) = i$ where $f_i=||F_i||$, the differential
\begin{equation}
    df_i = \sum_j\frac{\partial f_i}{\partial a_j}da_j
\end{equation}
thus contains the sensitivity of level feeding from different feeding histories. This is in contrast to the work in \cite{athanasios}, where partial variance may now by determined for total cascades that feed the level, not individual transitions:
\begin{equation}
    \Delta\text{Var}_{ij} = \frac{\partial f_i}{\partial a_j}\sigma^2_j
\end{equation}
where $j$ denotes a path to level $i$. This adds further detail to the analysis of which transitions uncertainties are important because it offers through which cascade that importance is established. The work in \cite{athanasios} emphasizes the ability to state which transitions need further measurement for improved calculations. The path algebra allows one to state which cascade is most responsible for that transitions importance thus establishing relational importance to other transitions. 
\subsection{Conclusion}

Motivated by the need to extend the generality of the transition matrix formalism traditionally used to model nuclear decay schemes, We have demonstrated how the path algebra of a decay quiver gives the underlying algebra for transitions matrices. This allowed for an extension into a tensor decay space which we have restricted by defining a coincidence vector space upon which we build the coincidence algebra. The coincidence algebra allows for a more general algebra for calculating coincidence probabilities as it allows for multiplication between non connected paths in the decay quiver and in addition separates different branching terms. This is valuable as it allows for the prediction of a complete full energy summing spectrum between all transitions--not merely composable. With the path algebra being the space with which nuclear levels schemes live, we have defined an algebra bundle over the path algebra where each decay vector defines the structure constants on the coincidence algebra at the fiber. We have showed, at the very least that this algebra allows for additional coincidence probability terms to be calculated to aid in the predictions for energy spectra in gamma-ray spectroscopy. The path algebra framework also adds interesting detail to the discussion of uncertainty propagation in nuclear level schemes. 

\section*{Software and Data Availability}
The \texttt{COINAlgebra} software used for creating nuclear levels and their analysis may be available upon request, though as stated it is still in its early development stages as a general software for distribution.

\section*{Acknowledgments}
 This work was supported by the Natural Sciences and Engineering Research Council of Canada (NSERC).

\newpage 

\bibliographystyle{elsarticle-num} 
\bibliography{biblio}
\end{document}